\documentclass[aps,pre,twocolumn,superscriptaddress,showpacs]{revtex4}
\usepackage{graphicx}
\usepackage{times}
\begin{document}
\def\T{\Theta}
\def\D{\Delta}
\def\d{\delta}
\def\r{\rho}
\def\p{\pi}
\def\a{\alpha}
\def\g{\gamma}
\def\ra{\rightarrow}
\def\s{\sigma}
\def\b{\beta}
\def\e{\epsilon}
\def\G{\Gamma}
\def\om{\omega}
\def\pe{$1/r^\a$ }
\def\l{\lambda}
\def\f{\phi}
\def\w{\psi}
\def\m{\mu}
\def\t{\tau}
\def\dg{d^3{\bf r}\,d^3{\bf p}}
\def\df{f({\bf r, p})}
\def\dn{n({\bf r, p})}

\title{Anomalously slow phase transitions in self-gravitating systems}
\author{I.~Ispolatov}
\affiliation{Departamento de Fisica, Universidad de Santiago de Chile,
Casilla 302, Correo 2, Santiago, Chile}

\author{M.~Karttunen}
\affiliation{Biophysics and Statistical Mechanics Group, 
Laboratory of Computational Engineering, Helsinki University
of Technology, P.\,O. Box 9203, FIN-02015 HUT, Finland}

\date{\today}

\begin{abstract}
Kinetics of collapse and  explosion transitions 
in microcanonical self-gravitating ensembles is analyzed.   
A system of point particles interacting 
via an attractive soft Coulomb potential 
and confined to a spherical container is considered. We observed that 
for 100--200 particles  collapse 
takes $10^3$ -- $10^4$ particle
crossing times to complete, i.\,e., it is  by 2-3 orders of
magnitude slower than velocity relaxation. In addition, it is found
that the collapse time decreases rapidly with the increase of 
the softcore radius. We found that such an anomalously
long collapse time is caused by the slow energy exchange between a
higher-temperature compact core and relatively cold diluted halo.
The rate of energy exchange between the faster modes of the
core particles and slower-moving particles of the halo
is exponentially small in the ratio of the frequencies of these modes.
As the softcore radius increases, 
and the typical core modes become slower,  the ratio
of core and halo frequencies decreases and the collapse accelerates.
Implications to astrophysical systems and phase transition 
kinetics are discussed.
 
\end{abstract}

\maketitle
\section{Introduction}
\label{sec_intro}

Many groups of stellar systems have highly universal structures
despite the apparent differences in their history and 
environment~\cite{bt, astro}.  
These universal structures are thought to have arisen as a result
of relaxation towards
equilibrium or to otherwise long-lived states. 
A comparison between the age of the stellar systems
with universal features
and corresponding collisional relaxation times 
reveals, however, that several types of stellar systems,
such as elliptical galaxies, have not existed long enough to be 
collisionally relaxed~\cite{bt}.
Other than collisional types of relaxation, such as ''violent
relaxation'', or phase-space mixing caused by strong gravitational field
fluctuations~\cite{lb},  have been suggested to explain this apparent
contradiction 
between the timescales. Yet a
full understanding of the kinetics of relaxation in naturally occurring
self-gravitating systems systems is still lacking. 

A number of fairly idealized models have been analyzed to understand the
nature of equilibrium  and transitory states of  stellar systems.
A well-studied example  is
an ensemble of self-gravitating particles 
particles with a sufficiently 
short-range small distance regularization confined in a container.
That system exhibits a gravitational phase transition 
between a relatively uniform
high-energy state and a low-energy state with a core-halo 
structure~\cite{pr,ki2,ch1,bm2,usg,dv,ch,chs,chi,grv,pet,jm}.
During such a transition in a microcanonical ensemble
the system undergoes a discontinuous jump
from a state that just ceases to be a local entropy maximum to a global
entropy maximum state with the same energy but different temperature.
A transition from a high-energy uniform state 
to a lower-energy core-halo state is usually called collapse.
The reverse transition during which the core disappears
is often referred to as explosion.

It has been recently observed in molecular dynamics (MD) simulations 
\cite{jm} that a typical timescale for
such gravitational transitions is paradoxically large, for a system
of 125 -- 250 particles being of the order of
$10^3$  relaxation times for a collapse and 
$10^2$ relaxtion times for an explosion. The relaxation
time,  $t_r={R^{3/2} N^{1/2}}/{\ln N}$, is the timescale of  typical particle
velocity thermalization which proceeds mostly via soft Coulomb 
collisions~\cite{bt}. 
It was also observed in Ref.~\cite{jm} that the density relaxation, such as
the formation of the core, advances relatively fast, while the
evolution of the kinetic energy or temperature  proceeds
noticeably slower~\cite{jm}. 

In this paper we undertake a more detailed study of kinetics of
collapses in self-gravitating systems, in some way
completing the investigation initiated in~\cite{jm}. 
The structure of the paper
is following: After this introduction we briefly outline the simulation 
setup and present the results for collapse kinetics. A section 
analyzing a slow core-halo energy transfer as the bottleneck of system
relaxation follows. Conclusions and
discussion of the results completes the paper. 

\section{simulations}

We consider systems consisting of $N=125-250$ identical particles of unit
mass confined in a spherical container of radius $R$ with reflecting walls.
The Hamiltonian of the system reads as
\begin{equation}
\label{ham}
H=\sum_{i=1}^N \frac{p_i^2}{2} - \sum_{i<j}^N\frac{1}{\sqrt{r_{ij}^2+r_0^2}},
\end{equation}
where $r_0$ is the soft core radius.
Along with the physical units, we use the standard rescaled
units (as discussed, e.\,g., in Ref.~\cite{pr}).
For  energy $\e$, temperature $\T$, distance $x$, 
and time $\t$ they read as 
\begin{eqnarray}
\label{def}
\nonumber
\e & \equiv & E{R\over N^2}\\
\nonumber
\T & \equiv & T {R\over N}\\
x & \equiv & {r\over R}\\
\nonumber
\t & \equiv & t \sqrt{N \over R^3}.
\end{eqnarray}
Expressed in these rescaled units, 
equilibrium properties of self-gravitating systems become universal.
The velocity relaxation, assuming that it is caused mostly by the 
soft collisions, is expected to be universal in terms of time
$\t_{vel}=\t \ln N /N$~\cite{bt}, where the factor $ N/\ln N$ is proportional
to 
the number of crossings a particle needs to change its velocity by a factor of
two.

\begin{figure}
\includegraphics[width=.45\textwidth]{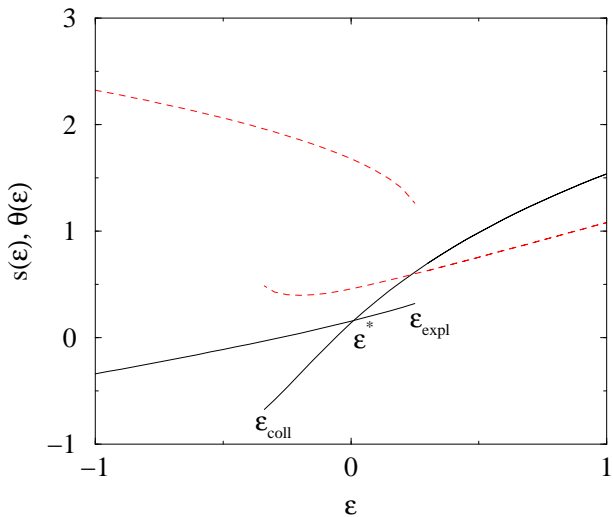}
\caption{\label{fig_mf}
Plots of entropy $s(\e)$ (solid line) and 
temperature $\t(\e)=d\e/ds$ (dashed line) vs energy $\e$
for a system with a gravitational phase transition and a softcore radius
$x_0=0.005$.}
\end{figure}

The phase diagram of the system is presented in Fig.~\ref{fig_mf}, see also
Refs.~\cite{usg,chi,jm}. 
High- and low-energy branches terminating at the
energies $\e_{coll}$ and $\e_{expl}$ correspond to the uniform and core-halo
states.  The collapse and
explosion energies
are $\e_{coll}\approx -0.339$ and $\e_{expl}\approx 0.267$ for 
$x_0=5\times 10^{-3}$. 

Each MD run was initiated with a configuration in which
the particles were seeded according to the corresponding $\e_{coll}$ or 
$\e_{expl}$
equilibrium (metastable) density profiles, and
the velocities were assigned according to the Maxwell distribution.
A more detailed description of the MD simulation procedure is presented
in~\cite{jm}. 
  
To reveal all facets of
gravitational phase transitions in the most informative way, we consider the
following parameters:
\begin{itemize}
\item Temperature $\T$ as the indicator of the advancement
of a phase transition as a whole. When  $\T$
reaches the target phase equilibrium value, all other system parameters
come to equilibrium as well and the phase transition is complete.
\item Number of core particles $N_{c}$ as the measure of  density relaxation.
\item Temperature of the core $\T_{c}$ which is proportional to the average 
kinetic energy of the core particles.
Deviations of $\T_{c}$ from  $\T$ quantify the temperature
gradients occurring  during a phase transition.
\end{itemize}

We do not list here any parameters which characterize the velocity relaxation:
As it follows from the definition of $\t_{vel}$~\cite{bt}, 
and as observed in simulations~\cite{jm}, the 
velocity distributions in both core and halo become thermalized within 
$\t_{vel} \sim 1$.

\begin{figure}
\includegraphics[width=.45\textwidth]{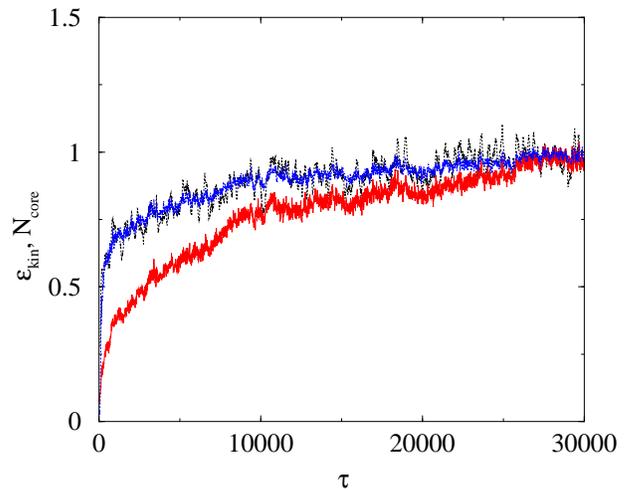}
\caption{\label{fig_ec}
Plots of the relative values of
(from top to the bottom) the number of core particles $N'_{c}(\t)$ (blue),
core temperature $\T_{c}(\t)$ (almost coinciding with $N'_{c}(\t)$, but with
stronger fluctuations, dotted black), and total temperature  $\T'(\t)$ (red)
vs $\t$
for a collapse in system with $\e=-0.5$, $N=125$, and $x_0=0.005$}
\end{figure}

Results, averaged over 4 runs, for the temperature, core temperature and the
number of core particles  for a collapsing system are presented in  
Fig.~\ref{fig_ec}. Time evolution of these parameters is plotted   
in terms of relative variables  $\T'(\t)$, $\T_{c}'(\t)$, and
$N_{c}'(\t)$ which 
are defined as 
$\T'(\t)\equiv[\T(\t)-\T(u)]/[\T(c-h)-\T(u)]$.
The values  $\T(u)$ and $\T(c-h)$ correspond to
the uniform and core-halo states in equilibrium. 

As in Ref.~\cite{jm}, we observe that a collapse in a system with
$N=125-250$ particles and $x_0=0.005$ takes about $10^3$ velocity
relaxation times to 
complete.  It also follows from Fig.~\ref{fig_ec}  that the growth of the 
core is significantly faster than the relaxation of the average kinetic
energy: The core reaches half of its equilibrium size only in 
about 5 velocity
relaxation times,  while temperature relaxes to half-way only
in 110 velocity relaxation times.
In addition, we can conclude that the core temperature is evolving 
almost synchronously with the number of core particles, i.\,e., noticeably faster
than the total temperature of the system. 
\begin{figure}
\includegraphics[width=.45\textwidth]{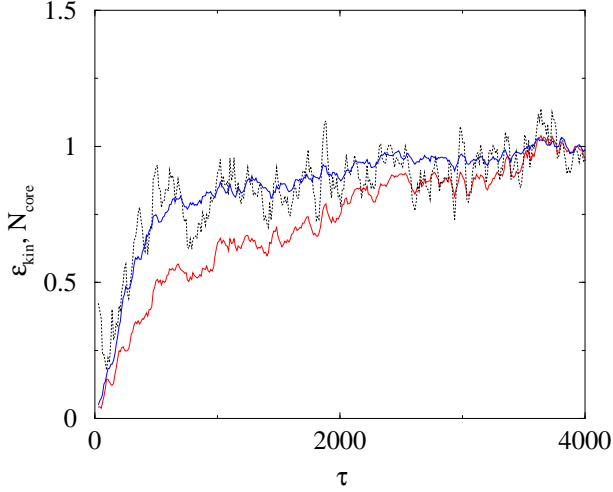}
\caption{\label{fig_ec1}
Same as in Fig.~\ref{fig_ec} but for $x_0=0.01$}
\end{figure}

The results for the collapse in an otherwise identical system but with softcore radius
twice larger, $x_0=0.01$, are presented in Fig.~\ref{fig_ec1}.
It follows from the comparison between  Figs.~\ref{fig_ec} and  \ref{fig_ec1}
that while the initial stages of relaxation are not affected by the
change of the short-range potential, the overall collapse proceeds much faster
for larger $x_0$.

The above numerical results suggest that after
a rapid initial core growth, which takes few velocity relaxation
times, a further evolution is hindered by some
slow process that is  essential for the completion of the phase transition.
Kinetics of this slow process strongly depends on the short-range part
of the interparticle potential.
It follows from Figs.~\ref{fig_ec} and \ref{fig_ec1} that a
collapse comes to its completion only when the temperatures of the core and the halo
become equal. Hence, it seems natural to assume that the bottleneck 
process is the energy, or heat, exchange between the core and the rest 
of the system. The rate of this heat exchange depends on the
structure of the core which in turn is determined by the potential
softening.
In the next section we consider the heat exchange between the core
and halo in more detail. 

\section{Core -- halo energy exchange}

To analyze the energy exchange between the core and halo, let us first 
examine the motion of a core particle.
In a system of reference of the center of mass (CM) of the core we expand the
potential energy terms in the Hamiltonian, Eq.~(\ref{ham}), in  powers of
$r_{ij}/r_0$ and arrive to the harmonic oscillator Hamiltonian,
\begin{equation}
\label{hamcore}
H\approx -{N_{c}\over 2 r_0^3} + {1 \over 2} \sum_{i=1}^{N_c} p_i^2 +
{N_{c}\over r_0^3} r_i^2. 
\end{equation}
Hence the motion of core particles relative to the core CM is characterized
by harmonic oscillations  with the frequency 
\begin{equation}
\label{fc}
\om_{c}= \sqrt{N_{c} \over r_0^3}.
\end{equation}

It is interesting to note that the frequency of the motion of a particle
in a uniform
self-gravitating sphere of $N_{c}$ particles and radius $r_0$
with the bare gravitational ($-1/r$) interaction is also given by
Eq.~(\ref{fc}). Since the core radius is roughly equal to $r_0$~\cite{jm},
both the bare interaction and the ''very soft'' potential 
frequencies are essentially the same.
\begin{figure}
\includegraphics[width=.45\textwidth]{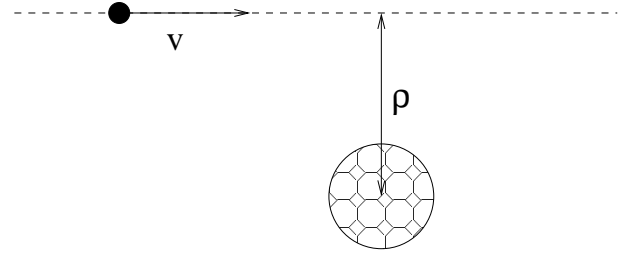}
\caption{\label{fig_sk}
Sketch of a core -- halo particle scattering event.}
\end{figure}

While the motion of the core particles is bound by gravity, the higher energy
halo particles
can be viewed as free and being confined by the container walls only. 
Hence, the interaction between a core and a halo particle can be 
approximated as an interaction between a point mass on a rectilinear
trajectory and a harmonic oscillator. The first relevant term  in the multipole
expansion of this 
interaction is the monopole-dipole term
\begin{equation}
\label{V}
V(t)=\sum_{i=1}^{N_c}\frac{x_i vt+z_i \rho}{[(vt)^2+\rho^2]^{3/2}}.
\end{equation}
Here $x_i$ and $z_i$ are the coordinates of core particles, $v$ and $\rho$ are
the velocity and the impact parameter of the halo particle, respectively (Fig.~\ref{fig_sk}). 
The monopole-monopole term, 
corresponding to the interaction between the halo particle and the core CM, is
irrelevant to the internal motion of the core particles.

With the introduced approximations  the core--halo energy exchange 
becomes physically identical to the well-studied process
of energy exchange between the vibrational
degrees of freedom of bound states and fast free particles in a plasma.
Naturally, such molecular processes are usually considered in
quantum mechanical terms. Following a standard textbook~\cite{ns}
we start with the first-order term of the interaction representation expansion
of the perturbation of the oscillator wave function during a complete single
collision, 
\begin{equation}
\label{c}
C=-\frac{i}{\hbar}\int_{-\infty}^{+\infty}V(t)\exp(i\omega_{c} t) dt
\end{equation}
and arrive to the following expression for the probability $|C|^2$ of the
excitation from
(or de-excitation to) the ground state of a harmonic oscillator, 
\begin{equation}
\label{hamcoll}
|C|^2=\frac{2\omega_{c}^2 N_{c}^2}{\hbar v^4}\left[K_0^2\left(\frac{\om_{c}
  \rho}{v}\right)+ 
K_1^2\left(\frac{\om_{c} \rho}{v}\right)\right].
\end{equation}
Here $K_j$ are the McDonald
functions (Modified Bessel functions of the Second Kind), the factor
$N_{c}^2$ appears since the interaction potential
in Eq.~(\ref{hamcoll}) 
is the sum of $N_{c}$ identical terms and the probability is quadratic
in $V(t)$. Multiplying Eq.~(\ref{hamcoll}) by the transferred energy 
$\hbar\omega_c$, 
we get rid of the quantum constants and obtain an
expression for the typical energy exchange during
a collision between a core and a halo particle,
\begin{eqnarray}
\label{de}
\nonumber
\delta E(\rho,v)=\frac{2\omega_{c}^2 N_{c}^2}{
  v^4}\left[K_0^2\left(\frac{\om_{c} 
  \rho}{v}\right)+ 
K_1^2\left(\frac{\om_{c} \rho}{v}\right)\right]\\
\mathop
{\longrightarrow}_{{\om_{c} \rho/v} \to \infty} \frac {2\pi \om_{c}
N_{c}^2} 
{v^3 \rho}\exp\left(-\frac{2 \om_{c} \rho}{v}\right).
\end{eqnarray}
The same expression can be obtained by completely classical analysis 
considering the
energy transfer during forced oscillation~\cite{ll}.
The last limit in Eq.~(\ref{de}) is taken since $\rho/v$, which is of order of a
halo particle 
crossing time $R/v$, is much larger than a period of oscillation of the core
particle $1/ \om_{c}$. It follows from  Eq.~(\ref{de}) that the rate of energy
exchange between the core and halo particles is 
exponentially small in the ratio of typical frequencies of their motion.

To obtain the rate of energy transfer between the core and all halo particles 
per unit time, one needs to average
Eq.~(\ref{de}) over impact parameters and velocities,
\begin{equation}
\label{dedt}
\frac {\Delta E}{\Delta t}=\int_{r_0}^R 2\pi \rho d\rho 
\int v W_M(v)  n \delta E(\rho,v) d^3 v.
\end{equation}
Here the lower limit of integration for the impact parameter is set equal to
the core radius for
the dipole approximation to be correct, $W_M(v)$ is the Maxwell distribution,
$n=3N/(4\pi R^3)$ is the particle density. Using that $r_0 \ll R$ and
evaluating the velocity integral 
in the steepest descent approximation, the following expression is obtained
\begin{equation}
\label{dedtf}
\frac {\Delta E}{\Delta t} \approx \frac{\pi \sqrt{3} N_{c}^2 N (2 T
  \omega_{c} 
r_0)^{1/3}} {T R^3} \exp \left[-3 \left(\frac{\omega^2
  r_0^2}{2T}\right)^{1/3}\right]. 
\end{equation}
Assuming that the typical time of core -- halo relaxation $t_{c-h}$ 
is roughly equal to the ratio of the total  transferred energy
to the rate of the transfer and using Eq.~(\ref{fc}), 
we arrive to the following expression for $t_{c-h}$,
\begin{equation}
\label{tr}
t_{c-h}=\sqrt{\frac {R^3}{N}} \Delta \e \frac{\theta^{2/3}(x_0/n_{c})^{1/6}}
{\pi \sqrt{3} n_{c}^2} \exp  \left [-3 \left(\frac{n_{c}}
    {2x_0\theta}\right)^{1/3}\right]. 
\end{equation}
Here $ \Delta \e$ is the total  transferred energy in rescaled units and
$n_{c}=N_{c}/N$.

To evaluate the numerical value of $t_{c-h}$ we consider the 
example from Fig.~\ref{fig_ec} with the following parameters: $\e=-0.5$,
$x_0=0.005$, $n_{c}\approx 0.2$, $\theta\approx(\e_{kin}^{c-h}
- \e_{kin}^{u})/3\approx 1.2$ (which is the average between the initial and
final halo temperatures)~\cite{jm}. 
We assume that the total change in kinetic
energy of the system during the collapse is a result of the energy transfer from 
the core and set $\Delta \e \approx \e_{kin}^{c-h}
- \e_{kin}^{u} \approx 3.6$. With these parameters we obtain 
\begin{equation}
\label{trn}
\t_{c-h}\equiv t_{c-h}\sqrt{\frac {N}{R^3}}\approx 16000
\end{equation}
for the relaxation time.
Given the number of approximations used in obtaining Eq.~(\ref{tr}), the agreement
with the simulational result for the complete phase transition time 
$\t_{c-h}^{MD}\approx 27000$  (see Fig.~\ref{fig_ec}) is surprisingly good.
The agreement is even better for systems with the same energy  $\e=-0.5$ 
but larger softcore radius, $x_0=0.01$, where  $n_{c}\approx 0.22$,
$\theta\approx 0.83$,  and $ \Delta \e\approx 1.5$. The theoretical estimate,
Eq.~(\ref{tr}), yields $\t_{c-h}\approx 3600$ which is only very little below the simulational
result $\t_{c-h}^{MD}\approx 3800$.

The most significant contributions to underestimating the value of  $\t_{c-h}$ are
the following:
\begin{itemize}
\item All collision were considered complete, i.\,e., the integral in Eq.~(\ref{c}) had
   infinite limits. This is certainly not true for a confined system 
   especially for large impact parameters.
\item 
   The energy exchange between the core oscillations and halo particles was
   always considered in one direction, i.\,e., from core to halo. In reality,
   this is only true for slow halo particles. Close to equilibrium, the exchange
   becomes progressively mutual with increasing number of fast halo
   particles losing their energy to the core vibrations.
\end{itemize}

Other factors not taken into account in our estimate, but possibly affecting
the core-halo energy relaxation are: Quadrupole and higher-order terms in
the potential expansion, overlapping collision, more complex dynamics 
than oscillation and rectilinear motion of the core and halo particles, higher than
one photon processes, or higher order terms in the perturbation expansion
to mention some. Yet we believe that our relatively simple approach presented
above captures the essence of the core-halo energy relaxation and 
has semi-quantitative predictive power. 

\section{discussion and conclusion}
In the previous two sections we obtained the following results for the
kinetics of collapse from the uniform  to the core-halo state 
in self-gravitating systems:
\newcounter{ref}
\begin{list}{\arabic{ref}.}{\usecounter{ref} \itemsep=0pt \leftmargin=1cm}
\item The molecular dynamics simulations revealed that in a system of 100-200
particles the collapse time is
by two-three order of magnitude longer than the velocity relaxation 
and strongly depends on the short-range part of the interaction potential. 
\item 
It was found that the non-equilibrium feature with the slowest
relaxation time is the temperature difference between the core and the halo. 
In contrast, such parameters as number of core particles and core kinetic 
energy evolve relatively fast.
\item A mechanism similar to the vibrational--translational relaxation in
plasmas was suggested for the core -- halo energy exchange.
For this mechanism we show that the core-halo thermalization is 
exponentially slow in the ratio of typical frequencies of the motion of
core and halo particles. 
Despite several rather strong approximations used in our analysis, 
a theoretical estimate for the relaxation time is in a good agreement 
(less than by a factor of 2 off) with the simulation results.
\end{list}

\begin{figure}
\includegraphics[width=.45\textwidth]{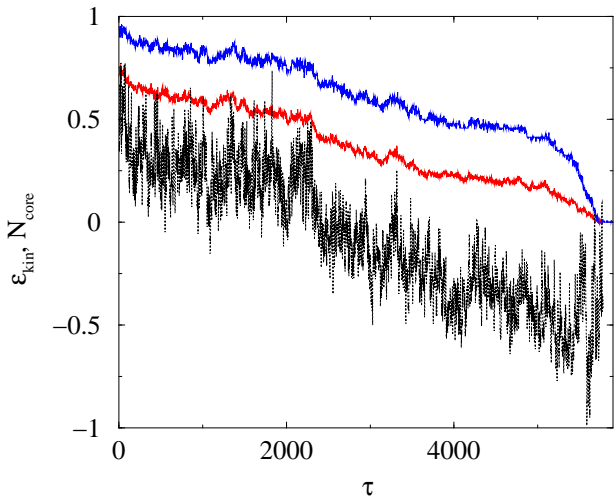}
\caption{\label{fig_ec2}
Plots of the relative values of
(from top to the bottom) number of core particles $N'_{c}(\t)$
(blue), total temperature  $\T'(\t)$ (red), and core
core temperature $\T_{c}(\t)$ (black)
vs $\t$
for the explosion in system with $\e=0.5$, $N=125$, and $x_0=0.005$}
\end{figure}

So far nothing has been said about the reverse to collapse transition, i.\,e.,
explosion, illustrated in Fig.~\ref{fig_ec2}.
It follows from this Figure that similarly to the collapse,
the kinetic energy is the slowest-evolving quantity
while the number of core particles and core kinetic energy lead the explosion.
Since the fastest particles leave the core first, the core is always colder
than the rest of the system. At the end of explosion the core becomes just a
single cold particle  indistinguishable from any other
particle of the system. This explains the sudden jump of the core temperature 
at the very end of explosion. Using Eq.~(\ref{tr}) it is straightforward to
explain why an explosion is faster than a collapse: Since the explosion starts
at higher energy than collapse, the corresponding 'explosion core-halo state'
has noticeably fewer core particles (twice for the considered here case) than
the final collapse state.
This reduces the exponential term in  Eq.~(\ref{tr}). In addition, the
total amount of energy which needs to be transferred between core and halo
is smaller in case of explosion than in the case of collapse. It follows from
Fig.~\ref{fig_mf} where temperature jumps at collapse and explosion points
are proportional to the energy exchanged between core and halo.

Are these results applicable to self-gravitating systems
with other types of short-range regularization?
For systems with
continuous potentials at low-energies the 
particle motion near the equilibrium position are
harmonic oscillations with a frequency roughly given by Eq.~(\ref{fc}) where
$r_0$ is of order of the core radius. Examples include
ensembles with Fourier-truncated Coulomb potentials~\cite{pet}
or exchange interaction in systems with
phase-space exclusion~\cite {ch1}. For potentials with singular short-range
part such as hard-core repulsion, the motion of core particles is
discontinuous, yet a typical inverse timescale of such motion is given
by  Eq.~(\ref{fc}) as well. This follows from the fact that the expression
Eq.~(\ref{fc}) 
can be obtained by dividing a typical core particle velocity $\sqrt{N_c/r_0}$
by the core radius $r_0$. The analysis and conclusion made in the Section III
are based on a general principle that for a perturbation of a fast system by
a slow one, the rate of energy transfer between these system is exponentially
small in the ratio of their frequencies. And since the motion of core
particles is always faster than the motion of the halo ones, 
we conclude that the results obtained here are qualitatively 
applicable to all collapsing self-gravitating
systems independent of the form of the short-range cutoff.

An important implication of the demonstrated here
essential non-isothermicity of collapsing system is about the
applicability of the
Smoluchowski equation description to these systems~\cite{chs}.
Compared to the
more complete Boltzmann equation with Landau collision term
(often called in case of self-gravitating systems Fokker-Plank-Vlasov 
equation~\cite{ki3}), Smoluchowski equation offers a significant
simplification: 
In the physically relevant case of spherically symmetric systems the
solution depends only on one radial coordinate $r$. 
To make such description more realistic, one needs to take into account the
non-uniform temperature field. This can be achieved by coupling the
Smoluchowski and heat conduction equations.
The non-isothermicity
of evolution also indicates that the collapse in canonical ensemble
must be radically different from its  microcanonical counterpart.

Finally few words about the astrophysical relevance of the obtained results.
The structure and mere existence of the equilibrium core is a consequence of 
short-range regularization of the gravitational potential and the
confining container which are the artifacts of the model. 
Hence the true equilibrium
core-halo states never occurs in stellar systems. However, the observation
that the collapse progress in hindered by the core  -- halo energy exchange
which is exponentially slow in the ratio of the typical frequencies
of motion of the core and halo particles remains applicable. 
Hence it is possible to interpret the states of the systems such as globular
clusters whose age is noticeably larger than corresponding velocity relaxation
(and therefore, initial core formation) time~\cite{bt} as the transitory
long-living states similar to core--halo thermalization states observed in
this study. 

\begin{acknowledgments}
The authors gratefully acknowledge the support
of Chilean FONDECYT under grants 1020052 and 7020052 
and the Academy of Finland grant No.~00119 (M.\,K.).
M.\,K. would like to thank the Department of Physics at 
Universidad de Santiago for warm hospitality.
\end{acknowledgments}

\end{document}